\documentstyle[12pt]{article}
\renewcommand{\theequation}{\arabic{equation}}
\newcommand{\EQ}{\begin{equation}}
\newcommand{\EN}{\end{equation}}

\newcommand{\ket}[1]{\left|#1\right\rangle}      % Ket-Zustand
\newcommand{\bear}{\begin{eqnarray}}
\newcommand{\ear}{\end{eqnarray}}

\begin{document}

\topmargin 0pt
\oddsidemargin 5mm
\newcommand{\NP}[1]{Nucl.\ Phys.\ {\bf #1}}
\newcommand{\PL}[1]{Phys.\ Lett.\ {\bf #1}}
\newcommand{\NC}[1]{Nuovo Cimento {\bf #1}}
\newcommand{\CMP}[1]{Comm.\ Math.\ Phys.\ {\bf #1}}
\newcommand{\PR}[1]{Phys.\ Rev.\ {\bf #1}}
\newcommand{\PRL}[1]{Phys.\ Rev.\ Lett.\ {\bf #1}}
\newcommand{\MPL}[1]{Mod.\ Phys.\ Lett.\ {\bf #1}}
\newcommand{\JETP}[1]{Sov.\ Phys.\ JETP {\bf #1}}
\newcommand{\TMP}[1]{Teor.\ Mat.\ Fiz.\ {\bf #1}}
     
\renewcommand{\thefootnote}{\fnsymbol{footnote}}
     
\newpage
\setcounter{page}{0}
\begin{titlepage}     
\begin{flushright}
UFSCARF-TH-96-10
\end{flushright}
\vspace{0.5cm}
\begin{center}
{\large  Algebraic Bethe ansatz approach for the one-dimensional 
Hubbard model
}\\
\vspace{1cm}
\vspace{1cm}
{\large  P.B. Ramos and M.J.  Martins } \\
\vspace{1cm}
{\em Universidade Federal de S\~ao Carlos\\
Departamento de F\'isica \\
C.P. 676, 13560~~S\~ao Carlos, Brasil}\\
\end{center}
\vspace{1.2cm}
     
\begin{abstract}
We formulate in terms of the quantum inverse scattering method the 
algebraic Bethe ansatz solution of the one-dimensional Hubbard model.
The method developed is based on a new set of commutation relations 
which encodes a hidden symmetry of 6-vertex type. 
\end{abstract}
\vspace{.2cm}
%\centerline{PACS numbers: 05.50+q, 64.60.Cn, 75.10.Hk, 75.10.Jm}
\vspace{.2cm}
%\centerline{May 1996}
\end{titlepage}

\renewcommand{\thefootnote}{\arabic{footnote}}

In 1968 the exact solution of the one-dimensional Hubbard 
model by the coordinate
Bethe ansatz was presented by Lieb and Wu \cite{Wu,SU} . It took some 
years until Shastry 
found the many conserved charges\cite{SA1} and 
also the two-dimensional classical vertex model\cite{SA2,SA3} 
whose transfer-matrix commutes with the Hubbard Hamiltonian. The
$R$-matrix responsible for the integrability(``infinite number of conserved charges'') was
then explicitly exhibited\cite{SA2,SA3}. Some time later, Wadati et al \cite{WA}
were able to verify such results by using a quite different 
and interesting approach. Afterwards, Bariev \cite{BA} developed a variant of
the coordinate Bethe ansatz to study  Shastry's vertex model, though on 
the basis of a diagonal-to-diagonal transfer matrix approach \cite{SO}.
More recently, some progress has been made concerning the Yangian symmetry 
of the Hubbard model \cite{KOU} and also on its ``free-fermion'' Yang-Baxter
structure \cite{JA}. Some further discussion about  Hubbard's invariant 
can also be found in the literature \cite{CC}.

However, certain important properties underlying such ``integrable'' program still needs
to be understood. This is justified, for example, by the early attempt of Shastry \cite{SA3}
in conjecturing the eigenvalues of the row-to-row 
transfer matrix  of the ``covering'' vertex 
model.  An important step towards closing this program is certainly the
formulation of the Bethe states of the one-dimensional Hubbard model  by
means of the quantum inverse scattering approach \cite{FAD}. Unlike the standard Bethe ansatz, this method
is based on first principle algebraic rules and definitely brings new insight on the mathematical
structure of integrable systems. The solution of the Hubbard model by the quantum inverse scattering
method is, in fact, a long-standing problem in the field of exactly solved models.
In this letter we show how
this more unified approach of Bethe ansatz technique can be established for the one-dimensional 
Hubbard model. In the course of our formulation we had to overcome a major
difficulty: the non-additive property of the Hubbard $R$ matrix. We have found the 
fundamental commutation rules between the creation and
annihilation operators present in the embedding vertex model. It turns out that the 
eigenvectors, eigenvalues and the Bethe ansatz equations follow as a consequence of  
systematic algebraic manipulation of such commutation rules. A hidden symmetry of $6$-
vertex type, important for  integrability, is noted.  We think that our results should also be of
relevance for future developments of the  physical properties of the one-dimensional Hubbard model. One
possibility should be the  application of our formulation in the context
of Korepin et al. method \cite{KO} of computing 
correlation functions. 

The Hamiltonian of the Hubbard model on a one-dimensional lattice of length $L$ is written in the form
\EQ
H=- \sum_{i=1}^{L} \sum_{\sigma=\uparrow,\downarrow}( c_{i, \sigma}^{\dagger} c_{i+1, \sigma} +h.c) 
+U \sum_{i=1}^{L} n_{i,\uparrow} n_{i,\downarrow}
\EN
where $c_{i,\sigma} (n_{i,\sigma}) $ are Fermi (number) operators with spin $\sigma$ on site $i$, and $U$ is the Hubbard coupling.
Fundamental to the integrability of the Hubbard model is the fact that Hamiltonian (1) commutes with certain one-parameter
family of transfer matrix $T$. In analogy with integrable systems in classical mechanics, $T$ is the generator of the many conserved
charges, and Hamiltonian (1) is one of those currents. Thus, the analysis of the physical properties of the transfer matrix will
certainly provide a deeper understanding of the Hubbard model. The appropriate two-dimensional classical statistical system
exhibiting such properties was found by Shastry \cite{SA2,SA3}. The model is constituted of two coupled $6$-vertex model satisfying the
free-fermion condition. The vertex model is parametrized in terms of 
three functions $a(\lambda)$,$b(\lambda)$ and $h(\lambda)$ which are constrained by Hubbard's coupling as
\EQ
sinh[2h(\lambda)] = \frac{U}{2} a(\lambda)b(\lambda)
\EN
where functions $a(\lambda)$ and $b(\lambda)$ are the non-trivial free-fermion Boltzmann weights. 
This gives us an $R$-matrix $R(\lambda,\mu)$ consisting of ten 
distinct Boltzmann weights. Here we denote them by
$\alpha_{i}(\lambda,\mu)$,$i=1, \cdots ,10$. In Appendix $A$ we
present the structure of the $R$-matrix, the explicit expressions and some
usefull identities for the 
weights $\alpha_i(\lambda,\mu)$ 
\cite{SA2,WA}.  In general, the transfer matrix $T$ is obtained as a trace of an auxiliary monodromy operator, $T=Tr_{G} \cal{T}$.
The space $G$ is a `` ghost '' variable, corresponding to a horizontal arrow in the classical vertex model. Its dimension corresponds
to the four possible states of the Hubbard model on a given site. 
As we shall see below, it is convenient to write the associated monodromy
matrix ${\cal T}(\lambda)$ as
\EQ
{\cal T}(\lambda) =
\pmatrix{
B(\lambda)       &   \vec{B}(\lambda)   &   F(\lambda)   \cr
\vec{C}(\lambda)  &  \hat{A}(\lambda)   &  \vec{B^{*}}(\lambda)   \cr
C(\lambda)  & \vec{C^{*}}(\lambda)  &  D(\lambda)  \cr}
\EN
where $\vec{B}(\lambda)$ $(\vec{B^{*}}(\lambda))$ and 
$\vec{C}(\lambda)$ $(\vec{C^{*}}(\lambda))$  are two component vectors 
with dimensions $1 \times 2$$(2 \times 1)$ and $2 \times 1$$(1 \times 2)$, respectively. The 
operator $\hat{A}(\lambda)$ is a $2 \times 2$ 
matrix and the other remaining operators are scalars. The integrability condition is based
on the Yang-Baxter algebra, namely
\EQ
R(\lambda , \mu) {\cal T}(\lambda)  \stackrel{s}{\otimes} {\cal T}(\mu) =
{\cal T}(\mu)  \stackrel{s}{\otimes} {\cal T}(\lambda)  R(\lambda ,\mu)
\EN
where the symbol $ \stackrel{s}{\otimes} $  stands for the Grassmann direct product \cite{KS}. Such definition takes
into account the extra signs appearing when fermionic states ( spin up and down ) are permuted \cite{WA}.
One consequence of  Shastry's Boltzmann
weights is that the monodromy matrix has a triangular form when acting on the standard
ferromagnetic pseudovacuum $\ket{0}$. More precisely, we find the following diagonal 
properties
\EQ
B(\lambda)\ket{0} = [\frac{a(\lambda)}{b(\lambda)}e^{2h(\lambda)}]^{L}\ket{0},~~ D(\lambda)\ket{0} = 
[\frac{b(\lambda)}{a(\lambda)}e^{2h(\lambda)}]^{L}\ket{0},~~ 
A_{aa}(\lambda)\ket{0} =\ket{0} , a=1,2
\EN
as well as the annihilation identities
\EQ
 C(\lambda) \ket{0} = \vec{C}(\lambda)\ket{0}=  
\vec{C^*}(\lambda)\ket{0} = 0,
 A_{ab}(\lambda)\ket{0} = 0 (a \neq b =1,2)
\EN

This suggests that operators  
$\vec{B}(\lambda)$,  $\vec{B^{*}}(\lambda)$ and $F(\lambda)$ act as creator fields on the 
ferromagnetic reference state $\ket{0}$. We notice, however, that the operators    
$\vec{B}(\lambda)$ and $\vec{B^{*}}(\lambda)$ do not mix under the 
integrability condition $(4)$. Therefore, in the construction of the eigenvectors it will be
enough to look only for combinations between the fields $\vec{B}(\lambda)$  and 
$F(\lambda)$. The one-particle state $\ket{\Phi_{1}(\lambda_{1})}$ is made by the linear
combination 
\EQ
\ket{\Phi_{1}(\lambda_{1})} = \vec{B}(\lambda_1). \vec{\cal{F}} \ket{0} = B_{a}(\lambda_{1}){\cal{F}}^{a}\ket{0}
\EN
where ${\cal{F}}^{a}$ is the component of a constant vector $\vec{\cal{F}}$ with dimension $(2 \times 1)$. The
two-particle state $\ket{\Phi_{2}(\lambda_{1},\lambda_{2})}$ depends both of operators 
$\vec{B}(\lambda)$ and $F(\lambda)$. This happens because the commutation rule between
two fields of type $\vec{B}(\lambda)$ generates the scalar operator $F(\lambda)$. This is
a constrain imposed by the integrability condition $(4)$, which reads
\EQ
\vec{B}(\lambda) \otimes \vec{B}(\mu) = \alpha_{1,2}(\lambda,\mu)
[ \vec{B}(\mu) \otimes \vec{B}(\lambda) ] .\hat{r}(\lambda,\mu)
%\nonumber \\
-i\alpha_{10,7}(\lambda,\mu)  
\{ F(\lambda)B(\mu) - F(\mu)B(\lambda) \} \vec{\xi}
\EN
where we define $\alpha_{a,b}(\lambda,\mu)=\alpha_{a}(\lambda,\mu)/\alpha_{b}(\lambda,\mu)$. The vector 
$\vec{\xi}$ and the matrix $\hat{r}(\lambda,\mu)$ have the following structures
\EQ
{\vec \xi} = 
\matrix{(
0  &1  &-1  &0 )  \cr},~
\hat{r}(\lambda,\mu) = 
\pmatrix{
1  &0  &0  &0  \cr
0  &a(\lambda,\mu)  &b(\lambda,\mu)  &0  \cr
0  &b(\lambda,\mu)  &a(\lambda,\mu)  &0  \cr
0  &0  &0  &1  \cr}
\EN
where
\EQ
b(\lambda,\mu) = \alpha_{8,1}(\lambda,\mu)\alpha_{9,7}(\lambda,\mu)
,~~ a(\lambda,\mu) + b(\lambda,\mu) = 1
\EN

Remarkable enough we have found that $\hat{r}$-matrix (9) is in fact factorizable. Moreover, when
properly parametrized, it has the same structure of that appearing in the isotropic 6-vertex
model.
We stress that such hidden
symmetry is crucial in our algebraic construction and plays a fundamental role on the
exact solution of the
Hubbard model.  In our opinion, this is the `` nice '' algebraic explanation for the fact that the
bare two-body scattering of the Hubbard Hamiltonian appears in the 6-vertex form \cite{Wu,SU}.
This result can be
established by performing the following change of variables
\EQ
\tilde{\lambda} = \frac{a(\lambda)}{b(\lambda)}e^{2h(\lambda)}
- \frac{b(\lambda)}{a(\lambda)}
e^{-2h(\lambda)} - \frac{U}{2}
\EN

By using the Boltzmann weights \cite{SA2,WA} ( see Appendix $A$ ) in equation (10) and by considering the new
variables defined in (11),  we are able to rewrite functions $a(\tilde{\lambda},\tilde{\mu})$, $b(\tilde{\lambda},\tilde{\mu})$ as
\EQ
a(\tilde{\lambda},\tilde{\mu}) = \frac{U}{\tilde{\mu}-\tilde{\lambda} + U},~
b(\tilde{\lambda},\tilde{\mu}) = \frac{\tilde{\mu}-\tilde{\lambda}}{\tilde{\mu}-\tilde{\lambda} + U}
\EN
which are precisely the non-trivial Boltzmann weights of the isotropic $6$-vertex model
\cite{FAD,DV,KO}.
Taking into account our considerations above, it is not difficult to check that the
two-particle state is given by
\EQ
\ket{ {\Phi}{_2}(\lambda_{1},\lambda_{2})} =\{ 
\vec {B}(\lambda_{1}) \otimes \vec{B}(\lambda_{2}) 
 +i\alpha_{10,7}(\lambda_{1},\lambda_{2}) 
F(\lambda_{1})( \vec {\xi} \otimes \vec {\Phi}_{0} ) B(\lambda_{2}) \} . \vec{\cal{F}} \ket{0}
\EN
where $\vec{\Phi}_{0}$ is the unitary constant. In fact, we have checked that all unwanted
terms generated by the eigenvalue problem can be canceled out through a unique Bethe
ansatz equation. Moreover, at least at this
level, the physical meaning of our construction is the following. While each 
component of the field $\vec{B}(\lambda)$ creates an electron with spin up or down, the
operator $F(\lambda)$ is responsible for the double occupancy on a given site of the lattice.
In general, the $n$-particle state can be constructed by induction and we have verified that
it satisfies the following recurrence relation
\EQ
\ket{\Phi_{n}(\lambda_{1}, \cdots ,\lambda_{n})} =  
\vec {\Phi}_{n}(\lambda_{1}, \cdots ,\lambda_{n}).\vec{\cal{F}} \ket{0}
\EN
where
\bear
\vec {\Phi}_{n}(\lambda_{1},\cdots ,\lambda_{n}) = 
\vec {B}(\lambda_{1}) \otimes \vec {\Phi}_{n-1}(\lambda_{2}, 
\cdots ,\lambda_{n})
 + 
%F(\lambda_{1})\vec{\xi} \otimes \sum_{j=2}^n 
\sum_{j=2}^n 
i\alpha_{10,7}(\lambda_{1},\lambda_{j})
\prod_{k=2,k \neq j}^{n} 
i\alpha_{2,9}(\lambda_{k},\lambda_{j}) 
\nonumber \\
%\prod_{k=2,k \neq j}^{n} 
%i\alpha_{2,9}(\lambda_{k},\lambda_{j})
\times \left [ \vec{\xi} \otimes F(\lambda_1) 
 \vec {\Phi}_{n-2}(\lambda_{2},\cdots,\lambda_{j-1},\lambda_{j+1},\cdots,\lambda_{n}) B(\lambda_j) \right ] \prod_{k=2}^{j-1} \alpha_{1,2}(\lambda_{k},\lambda_{j})
\hat{r}_{k,k+1}(\lambda_{k},\lambda_{j})
\ear

Let us now turn to the diagonalization problem.  The associated transfer matrix is obtained as a graded trace
of the monodromy matrix $\cal{T}(\lambda)$.  The graded structure takes into account the
fermionic degrees of freedom, and on the diagonal of $\cal{T}(\lambda)$ only $\hat{A}_{aa}(\lambda)$
contributes with a non-null Grassmann parity. Hence, the eigenvalue problem becomes
\EQ
[B(\lambda)-\sum_{a=1}^{2}A_{aa}(\lambda)+D(\lambda)]\ket{ {\Phi}_{n}(\lambda_{1},\cdots ,\lambda_{n})} =
\Lambda(\lambda,\{\lambda_{i}\}) \ket{ {\Phi}_{n}(\lambda_{1},\cdots ,\lambda_{n})}
\EN

In order to solve $(16)$ we  need the commutation rules between the diagonal and the creation operators. This is similar to
solving a problem of quantum mechanics on the Fock space, 
analogously to the role of the Heisenberg algebra  on the
solution of the harmonic oscillator. In our case, the necessary commutation relations can be obtained by an appropriate
manipulation of integrability condition (4). The procedure is rather cumbersome, and here we only list some
of them which are fundamental for further discussion. They are given by  
\bear
\hat{A}(\lambda) \otimes \vec{B}(\mu) = 
-i\alpha_{1,9}(\lambda,\mu)[\vec{B}(\mu) \otimes 
\hat{A}(\lambda) ]. \hat{r}(\lambda,\mu)
+i \alpha_{5,9}(\lambda,\mu) \vec{B}(\lambda) \otimes \hat{A}(\mu)   
\nonumber \\
\left \{ -i\alpha_{10,7}(\lambda,\mu) [\vec{B^{*}}(\lambda)B(\mu) 
+i\alpha_{5,9}(\lambda,\mu)F(\lambda)\vec{C}(\mu)  
%\right . \nonumber \\ \left .
-i \alpha_{2,9}(\lambda,\mu)F(\mu)\vec{C}(\lambda)] \right \} 
\otimes \vec{\xi}
\ear
\EQ
B(\lambda)\vec{B}(\mu) = 
i \alpha_{2,9}(\mu,\lambda) \vec{B}(\mu)B(\lambda) -i 
\alpha_{5,9}(\mu,\lambda) \vec{B}(\lambda)B(\mu),
\EN
\bear
D(\lambda)\vec{B}(\mu) = 
-i\alpha_{8,7}(\lambda,\mu) \vec{B}(\mu)D(\lambda) 
+ \alpha_{5,7}(\lambda,\mu) F(u)\vec{C^{*}}(\lambda)
 \nonumber \\ 
 - \alpha_{4,7}(\lambda,\mu) F(\lambda)\vec{C^{*}}(\mu)
- i\alpha_{10,7}(\lambda,\mu) 
\vec{\xi}. [ \vec{B^{*}}(\lambda) \otimes \hat{A}(\mu)] 
\ear

The eigenvalue $\Lambda(\lambda,\{\lambda_{i}\})$ can be calculated by keeping the
terms proportional to the eigenstate $\ket{ {\Phi}_{n}(\lambda_{1},\cdots ,\lambda_{n})}$. For 
example, by using several times the
first terms of the commutation relations $(17,18,19)$ we find the following structure
for eigenvalue $\Lambda(\lambda,\{\lambda_{i}\})$ 
\bear
\Lambda(\lambda,\{\lambda_{i}\}) = 
[\frac{a(\lambda)}{b(\lambda)}e^{2h(\lambda)}]^L 
\prod_{i=1}^{n} i \alpha_{2,9}(\lambda_{i},\lambda) + 
[\frac{b(\lambda)}{a(\lambda)}e^{2h(\lambda)}]^L 
\prod_{i=1}^{n} -i\alpha_{8,7}(\lambda,\lambda_{i})  
\nonumber \\
-\prod_{i=1}^{n} -i \alpha_{1,9}(\lambda,\lambda_{i}) 
\Lambda^{(1)}(\lambda,\{\lambda_{j}\},\{\mu_j\}) 
\ear
where $\Lambda^{(1)}(\lambda,\{\lambda_{i}\})$ is the eigenvalue of an auxiliary 
inhomogeneous problem related to the hidden $6$-vertex symmetry  we have mentioned
before. More precisely, such auxiliary problem is defined by
\EQ
\hat{r}_{b_{1}d_{1}}^{c_{1}a_{1}}(\lambda,\lambda_{1})
\hat{r}_{b_{2}c_{2}}^{d_{1}a_{2}}(\lambda,\lambda_{2}) \cdots
\hat{r}_{b_{n}c_{1}}^{d_{n-1}a_{n}}(\lambda,\lambda_{n})
{\cal{F}}^{a_{n} \cdots a_{1}} = \Lambda^{(1)}(\lambda,\{\lambda_{j}\},\{ \mu_j \})  
{\cal{F}}^{b_{n} \cdots b_{1}}
\EN

Fortunately such additional 
eigenvalue problem can be solved using the well known results of Faddeev et al \cite{FAD}.
New parameters $\{ \mu_j \} $ are then introduced in order to perform the diagonalization problem (21) .
Here we just have to adapt their algebraic results in order to consider the $6$-vertex
problem on an irregular lattice. Considering that this later eigenvalue problem has 
appeared in many different contexts in the literature \cite{FAD,DV,KO}, we just present
our final results. First it is  convenient to generalize a bit Shastry's parametrization
\cite{SA3} by introducing the new functions $z_{\pm}(x) $ as
\EQ
z_{-}(x) = \frac{a(x)}{b(x)}e^{2h(x)},~~ z_{+}(x)= \frac{b(x)}{a(x)}e^{2h(x)}
\EN

In terms of functions $z_{\pm}(x)$ and the variables $\{\tilde{\mu_j}\}$ introduced in (11), we find that the
eigenvalue (20) (modulo overall constant)can be written as 
\bear
\Lambda(\lambda,\{z_{\pm}(\lambda_{i})\},\{\tilde{\mu}_{j}\}) = 
[z_{-}(\lambda)]^L \prod_{i=1}^{n} \frac{b(\lambda)}{a(\lambda)} \frac{1+z_{-}(\lambda_{i})/z_{+}(\lambda)}{1-z_{-}
(\lambda_{i})/z_{-}(\lambda)} + [z_{+}(\lambda)]^L \prod_{i=1}^{n}  \frac{b(\lambda)}{a(\lambda)} \frac{1+z_{-}(\lambda_{i})z_{-}(\lambda)}{1-z_{-}(\lambda_{i})z_{+}(\lambda)} 
 \nonumber \\
-\prod_{i=1}^{n} \frac{b(\lambda)}{a(\lambda)} \frac{1+z_{-}(\lambda_{i})/z_{+}(\lambda)}{1-z_{-}(\lambda_{i})/z_{-}(\lambda)}
\prod_{j=1}^{m} \frac{z_{-}(\lambda)-1/z_{-}(\lambda) 
- \tilde{\mu}_{j} + U/2}{z_{-}(\lambda)-1/z_{-}(\lambda)
 - \tilde{\mu}_{j} - U/2}
\nonumber \\
-\prod_{i=1}^{n}  \frac{b(\lambda)}{a(\lambda)}
\frac{1+z_{-}(\lambda_{i})z_{-}(\lambda)}{1-z_{-}(\lambda_{i})z_{+}(\lambda)}
\prod_{j=1}^{m} \frac{1/z_{+}(\lambda)-z_{+}(\lambda) - \tilde{\mu}_{j} - U/2}{1/z_{+}(\lambda)-z_{+}(\lambda) - \tilde{\mu}_{j} + U/2}
\ear

Analogously, in order to cancel the unwanted terms, it is possible to show that the 
nested Bethe ansatz equations constraining the numbers $\{\lambda_{i}\}$,$\{\tilde{\mu}_{j}\}$ are then given by
\EQ
[z_{-}(\lambda_{k})]^L = 
 \prod_{j=1}^{m}  \frac{z_{-}(\lambda_{k})-1/z_{-}(\lambda_{k}) - \tilde{\mu}_{j} + U/2}{z_{-}(\lambda_{k})-1/z_{-}(\lambda_{k}) - \tilde{\mu}_{j} - U/2},
~k=1, \cdots, n
\EN
\EQ
\prod_{k=1}^{n}  \frac{z_{-}(\lambda_{k})-1/z_{-}(\lambda_{k}) - \tilde{\mu}_{l} - U/2}{z_{-}(\lambda_{k})-1/z_{-}(\lambda_{k}) - \tilde{\mu}_{l} + U/2} =
- \prod_{j=1}^{n} \frac{\tilde{\mu}_{l}-\tilde{\mu}_{j}+U}{\tilde{\mu}_{l}-\tilde{\mu}_{j}-U},~~l=1, \cdots, m 
\EN

To check the consistency of our results (23-25) one has to verify that $ \Lambda(\lambda,\{ z_{\pm}(\lambda_i) \}, \{ \tilde{\mu}_j \}) $
is free of poles for finite values of $\lambda$. In fact, the null residue condition on both direct ( $z_{-}(\lambda)$ ) 
and crossed  ($z_{+}(\lambda)$ ) poles lead us to 
the Bethe conditions (24,25). A possible physical application of the eigenvalue
result (23) is probably concerned with the finite temperature properties of
the one-dimensional Hubbard model \cite{KUB}. This is connected to the recent
developments of new powerful methods to deal with finite size effects
in integrable models \cite{KUP,CDV}. These techniques depend much on the diagonalization of the quantum transfer matrix (rather the one-dimensional Hamiltonian), a problem which we managed to  solve in this letter.
Lastly, it is also possible to rewrite the
nested Bethe ansatz equations (24,25) in terms of the original form presented by Lieb and Wu \cite{Wu}. In this case, one just 
needs to change $ \tilde{\mu}_j
\rightarrow 2i \tilde{\mu}_j $ and
relate the variables $\lambda_k$ with the lattice momenta $p_k$ \cite{SA3} by the relation
$z_{-}(\lambda_k) = e^{ip_k} $.

We would like to conclude this letter with the following comments. The eigenvalue $(23)$ is 
almost the one conjectured by Shastry in ref. \cite{SA3}. They differ by important phase 
factors, which are not easily obtained by using only phenomenological arguments. Our 
result (23-25) is connected with periodic boundary conditions, while that conjectured
by Shastry is related to a rather peculiar (sector dependent) toroidal boundary conditions.
The method we presented here is easily extended for a more general inhomogeneous model
 \cite{SA3,JA}. We expect that the only change in the Bethe ansatz equations (24,25) will
be on the terms proportional to the power of $L$. We plan to discuss these results in a more
detailed version of this letter \cite{MP}. Finally, some extra remarks are now
in order. It is possible to show, from the commutation rules between the
``dual'' field $\vec{B^{*}}(\lambda)$ and $F(\lambda)$, that a second 
$SU(2)$ $6$-vertex hidden symmetry is also present \cite{MP}. Thus, the two
$6$-vertex structure are tied up by the same field $F(\lambda)$. This resembles
much the constrain leading to the $SO(4)$ symmetry of the Hubbard chain \cite{SO4}. This is known to be of 
relevance for the Bethe ansatz completeness \cite{KOC}, for the classification of the elementary excitations \cite{KOS}, and can also
play an important role in the computation of correlation functions \cite{KO}.

\section*{Acknowledgements}
The work of P.B. Ramos is support by Fapesp. M.J. Martins is partially supported by Cnpq and Fapesp.

\centerline{\bf Appendix A :}
\setcounter{equation}{0}
\renewcommand{\theequation}{A.\arabic{equation}}
The structure of the $R$-matrix \cite{SA2,SA3,WA} is
{\scriptsize
\bear
 R(\lambda,\mu) = 
\pmatrix{
 \alpha_{2} &0 &0 &0 &0  &0  &0  &0  &0  &0  &0  &0  &0  &0  &0  &0    \cr
 0 &\alpha_{5} &0 &0 &-i\alpha_{9}  &0  &0  &0  &0  &0  &0  &0  &0  &0  &0  &0    \cr
 0 &0 &\alpha_{5} &0 &0  &0  &0  &0  &-i\alpha_{9}  &0  &0  &0  &0  &0  &0  &0    \cr
 0 &0 &0 &\alpha_{4} &0  &0  &-i\alpha_{10}  &0  &0  &i\alpha_{10}  &0  &0  &\alpha_{7}  &0  &0  &0    \cr
 0 &-i\alpha_{8} &0 &0 &\alpha_{5}  &0  
&0  &0  &0  &0  &0  &0  &0  &0  &0  &0    \cr
 0 &0 &0 &0 &0  &\alpha_{1}  &0  &0  &0  &0  &0  &0  &0  &0  &0  &0    \cr
 0 &0 &0 &i\alpha_{10} &0  &0  &\alpha_{3}  &0  &0  &-\alpha_{6}  &0  &0  &-i\alpha_{10}  &0  &0  &0    \cr
 0 &0 &0 &0 &0  &0  &0  &\alpha_{5}  &0  &0  &0  &0  &0  &-i\alpha_{8}  &0  &0    \cr
 0 &0 &-i\alpha_{8} &0 &0  &0  &0  &0  &\alpha_{5}  &0  &0  &0  &0  &0  &0  &0    \cr
 0 &0 &0 &-i\alpha_{10} &0  &0  &-\alpha_{6}  &0  &0  &\alpha_{3}  &0  &0  &i\alpha_{10}  &0  &0  &0    \cr
 0 &0 &0 &0 &0  &0  &0  &0  &0  &0  &\alpha_{1}  &0  &0  &0  &0  &0    \cr
 0 &0 &0 &0 &0  &0  &0  &0  &0  &0  &0  &\alpha_{5}  &0  &0  &-i\alpha_{8}  &0    \cr
 0 &0 &0 &\alpha_{7} &0  &0  &i\alpha_{10}  
&0  &0  &-i\alpha_{10}  &0  &0  
&\alpha_{4} &0  &0  &0    \cr
 0 &0 &0 &0 &0  &0  &0  &-i\alpha_{9}  &0  &0  &0  &0  &0  &\alpha_{5}  &0  &0    \cr
 0 &0 &0 &0 &0  &0  &0  &0  &0  &0  &0  &-i\alpha_{9}  &0  &0  &\alpha_{5}  &0    \cr
 0 &0 &0 &0 &0  &0  &0  &0  &0  &0  &0  &0  &0  &0  &0  &\alpha_{2}    \cr}
\nonumber \\
\ear
}
where the weights $\alpha_{i}(\lambda,\mu)$ ( normalized by $\alpha_5(\lambda,\mu)$ ) are given by
\EQ
%\alpha_{1}=
\alpha_{1}(\lambda,\mu)= 
\left\{e^{[h(\mu)-h(\lambda)]}a(\lambda)a(\mu)+e^{-[h(\mu)-h(\lambda)]}b(\lambda)b(\mu) \right \} \alpha_{5}(\lambda,\mu)
\EN
\EQ
%\alpha_{2}=
\alpha_{2}(\lambda,\mu)= 
\left \{ e^{-[h(\mu)-h(\lambda)]}a(\lambda)a(\mu)+e^{[h(\mu)-h(\lambda)]}b(\lambda)b(\mu) \right \} \alpha_{5}(\lambda,\mu)
\EN
\EQ
%\alpha_{3}=
\alpha_{3}(\lambda,\mu)=
\frac{ e^{[h(\mu)+h(\lambda)]}a(\lambda)b(\mu)
+e^{-[h(\mu)+h(\lambda)]}b(\lambda)a(\mu) }
{a(\lambda)b(\lambda)+a(\mu)b(\mu)} \left \{
\frac{\cosh[h(\mu)-h(\lambda)]}{\cosh[h(\mu)+h(\lambda)]} 
\right \} \alpha_{5}(\lambda,\mu)
\EN
\EQ
%\alpha_{4}=
\alpha_{4}(\lambda,\mu)=
\frac{e^{-[h(\mu)+h(\lambda)]}a(\lambda)b(\mu)+e^{[h(\mu)+h(\lambda)]}b(\lambda)a(\mu)}
{a(\lambda)b(\lambda)+a(\mu)b(\mu)}
\left \{ \frac{\cosh(h(\mu)-h(\lambda))}{\cosh(h(\mu)+h(\lambda))} 
\right \} \alpha_{5}(\lambda,\mu)
\EN
\EQ
%\alpha_{6}=
\alpha_{6}(\lambda,\mu)= \left \{
\frac{ e^{[h(\mu)+h(\lambda)]}a(\lambda)b(\mu)-e^{-[h(\mu)+h(\lambda)]}b(\lambda)a(\mu) }
{a(\lambda)b(\lambda)+a(\mu)b(\mu)} \right \} [b^{2}(\mu)-b^{2}(\lambda)]
\frac{\cosh[h(\mu)-h(\lambda)]}{\cosh[h(\mu)+h(\lambda)]} 
\alpha_{5}(\lambda,\mu)
\EN
\EQ
%\alpha_{7}=
\alpha_{7}(\lambda,\mu)= \left \{
\frac{-e^{-[h(\mu)+h(\lambda)]}a(\lambda)b(\mu)+e^{[h(\mu)+h(\lambda)]}b(\lambda)a(\mu)}
{a(\lambda)b(\lambda)+a(\mu)b(\mu)}
\right \} [b^{2}(\mu)-b^{2}(\lambda)]
\frac{\cosh[h(\mu)-h(\lambda)]}{\cosh[h(\mu)+h(\lambda)]} 
\alpha_{5}(\lambda,\mu)
\EN
\EQ
%\alpha_{8}=
\alpha_{8}(\lambda,\mu)= 
\left \{ e^{[h(\mu)-h(\lambda)]}a(\lambda)b(\mu)-
e^{-[h(\mu)-h(\lambda)]}b(\lambda)a(\mu) \right \} \alpha_{5}(\lambda,\mu)
\EN
\EQ
%\alpha_{9}=
\alpha_{9}(\lambda,\mu)= 
\left \{ -e^{-[h(\mu)-h(\lambda)]}a(\lambda)b(\mu)+
e^{[h(\mu)-h(\lambda)]}b(\lambda)a(\mu) \right \} 
\alpha_{5}(\lambda,\mu)
\EN
\EQ
%\alpha_{10}=
\alpha_{10}(\lambda,\mu)=
\frac{b^{2}(\mu)-b^{2}(\lambda)}{a(\lambda)b(\lambda)+a(\mu)b(\mu)}
\left \{ \frac{\cosh[h(\mu)-h(\lambda)]}{\cosh[h(\mu)+h(\lambda)]} 
\right \} \alpha_{5}(\lambda,\mu)
\EN

We remark that we have used the original Shastry's Boltzmann weights \cite{SA2,SA3} 
together with the grading  modifications of Wadati et al \cite{WA}. Moreover,
the 6-vertex parameters $a(\lambda)$ and $b(\lambda)$ satistfy the
free-fermion condition $a(\lambda)^2 + b(\lambda)^2 =1$. We also
list some important identities between the Boltzmann weights \cite{WA}
\EQ
\alpha_3(\lambda,\mu)= \alpha_1(\lambda,\mu) + \alpha_6(\lambda,\mu); 
\alpha_4(\lambda,\mu) + \alpha_7(\lambda,\mu)= \alpha_2(\lambda,\mu)
\EN
\EQ
\alpha_2(\lambda,\mu) \alpha_1(\lambda,\mu) -\alpha_9(\lambda,\mu) 
\alpha_8(\lambda,\mu)= \alpha_4(\lambda,\mu) \alpha_3(\lambda,\mu) -
\alpha_{10}^2(\lambda,\mu)= \alpha_5^2( \lambda, \mu)
\EN
\EQ
\alpha_2(\lambda,\mu) \alpha_3(\lambda,\mu) + \alpha_4(\lambda,\mu) 
\alpha_1(\lambda, \mu) = 2 \alpha_5^2(\lambda,\mu)
\EN

%\newpage

\end{document}